\documentclass[journal]{IEEEtran}

\ifCLASSINFOpdf
\else
\fi

\usepackage{url}
\usepackage[nolist]{acronym}
\usepackage{amsfonts}
\usepackage{xcolor}
\usepackage{amsmath}
\usepackage{graphicx}
\usepackage[normalem]{ulem}
\usepackage{footnote} 
\usepackage{bm}

\usepackage{amsmath}

\usepackage{url}
\usepackage{multirow}
\usepackage{amssymb}
\usepackage{tabularx} 
\usepackage{booktabs}
\usepackage{graphicx}    
\usepackage{subcaption}  
\usepackage{float}       

\begin{document}

\newacro{WER}[WER]{Word Error Rate}
\newacro{CER}[CER]{Character Error Rate}
\newacro{AVSR}[AVSR]{Audio-Visual Speech Recognition}
\newacro{ASR}[ASR]{Automatic Speech Recognition}
\newacro{CTC}[CTC]{Connectionist Temporal Classification}
\newacro{CE}[CE]{Cross Entropy}
\newacro{RNN-T}[RNN-T]{Recurrent Neural Network Transducer}
\newacro{SNR}[SNR]{Signal-to-noise ratio}
\newacro{AVSE}[AVSE]{Audio-Visual Speech Enhancement}
\newacro{MFCC}[MFCC]{Mel-Frequency Cepstral Coefficients}
\newacro{HMM}[HMM]{Hidden Markov Models}
\newacro{STFT}[STFT]{Short-time Fourier transform}
\newacro{MFCC}[MFCC]{Mel-Frequency Cepstral Coefficients}
\newacro{CNN}[CNN]{Convolutional Neural Network}
\newacro{LSTM}[LSTM]{Long Short-Term Memory}
\newacro{GRU}[GRU]{Gated Recurrent Unit}
\newacro{DNN}[DNN]{Deep Neural Network}
\newacro{HMM}[HMM]{Hidden Markov Model}
\newacro{RNN}[RNN]{Recurrent Neural Network}
\newacro{STOA}[SOTA]{State-of-the-Art}
\newacro{FPS}[FPS]{Frames per Second}
\newacro{FFN}[FFN]{Feed Forward Network}
\newacro{MHSA}[MHSA]{Multi-Head Self Attention}
\newacro{KL}[KL]{Kullback-Leibler}
\newacro{MTL}[MTL]{MultiTask Learning}
\newacro{AV-VAD}[AV-VAD]{Audio-Visual Voice Activity Detection}
\newacro{RBM}[RBM]{Restricted Boltzmann Machines}
\newacro{LM}[LM]{Language Model}
\newacro{PUnet}[P\&U net]{Predict-and-Update Network}
\newacro{DL}[DL]{Deep Learning}
\newacro{FE}[FE]{Factorize-Excitation}


\newcommand{\todo}[1]{\color{green} #1}
\newcommand{\etal}{\textit{et al.}}
\newcommand{\eg}{\textit{e.g.,}~}
\newcommand{\etc}{\textit{e.t.c.}}
\newcommand{\ie}{\textit{i.e.,}~}
\newcommand{\algs}{\textit{Algorithm}}

\newcommand{\fig}[1]{Fig.~}
\newcommand{\tab}[1]{Tab.~}
\newcommand{\Sec}[1]{Sec. }
\newcommand{\eq}[1]{Eq.}

\newcommand{\tofill}[1]{{\color{red}#1}}
\newcommand{\checkhere}[1]{{\color{green}#1}}
\newcommand{\qian}[1]{{\color{magenta}#1}}
\newcommand{\mat}[1]{{\textit{\textbf{}}#1}}
\renewcommand{\thefootnote}{}

\definecolor{burntumber}{rgb}{0.54, 0.2, 0.14}

\title{CueNet: Robust Audio-Visual Speaker Extraction through Cross-Modal Cue Mining and Interaction}


\author{
Jiadong Wang,~\IEEEmembership{Member,~IEEE}, 
Ke Zhang, 
Xinyuan Qian,~\IEEEmembership{Senior Member,~IEEE}, 
Ruijie Tao,
\\ Haizhou Li,~\IEEEmembership{Fellow,~IEEE}, 
Björn Schuller,~\IEEEmembership{Fellow,~IEEE}
\thanks{Jiadong Wang is with the University Hospital rechts der Isar, Technical University of Munich, Munich, Germany (e-mail: jiadong.wang@tum.de)}
\thanks{Ke Zhang is with the School of Artificial Intelligence, the Chinese University of Hong Kong (Shenzhen), China (e-mail: zhangke@cuhk.edu.cn)}
\thanks{Xinyuan Qian is with the Department of Computer Science and Technology, University of Science and Technology Beijing, Beijing, China. (email: qianxy@ustb.edu.cn)}
\thanks{Ruijie Tao is with the Department of Electrical and Computer Engineering, National University of Singapore, Singapore (e-mail: ruijie.tao@u.nus.edu)}
\thanks{Haizhou Li is with the Guangdong Provincial Key Laboratory of Big Data Computing, the Chinese University of Hong Kong (Shenzhen), China; also with  Shenzhen Research Institute of Big data, Shenzhen, China; also with the Department of Electrical and Computer Engineering, National University of Singapore, Singapore and also with the University of Bremen, Germany. (e-mail: haizhouli@cuhk.edu.cn)}
\thanks{Björn Schuller is with the University Hospital rechts der Isar, Technical University of Munich, Munich, Germany; also with Imperial College London, London, the United Kingdom. (e-mail: schuller@tum.de)}
}

\markboth{Journal of \LaTeX\ Class Files,~Vol.~14, No.~8, August~2022}%
{Shell \MakeLowercase{\textit{et al.}}: Bare Demo of IEEEtran.cls for IEEE Transactions on Magnetics Journals}

\IEEEtitleabstractindextext{%
\begin{abstract}
Audio-visual speaker extraction has attracted increasing attention, as it removes the need for pre-registered speech and leverages the visual modality as a complement to audio. Although existing methods have achieved impressive performance, the issue of degraded visual inputs has received relatively little attention, despite being common in real-world scenarios. Previous attempts to address this problem have mainly involved training with degraded visual data. However, visual degradation can occur in many unpredictable ways, making it impractical to simulate all possible cases during training. In this paper, we aim to enhance the robustness of audio-visual speaker extraction against impaired visual inputs without relying on degraded videos during training. Inspired by observations from human perceptual mechanisms, we propose an audio-visual learner that disentangles speaker information, acoustic synchronisation, and semantic synchronisation as distinct cues. Furthermore, we design a dedicated interaction module that effectively integrates these cues to provide a reliable guidance signal for speaker extraction. Extensive experiments demonstrate the strong robustness of the proposed model under various visual degradations and its clear superiority over existing methods.

\end{abstract}

\begin{IEEEkeywords}
Multimodality, speech extraction, feature disentanglement 
\end{IEEEkeywords}
}

\maketitle

\IEEEdisplaynontitleabstractindextext

\IEEEpeerreviewmaketitle

\section{Introduction}
\label{sec:intro}

\IEEEPARstart{A}{udio-Visual} Speaker Extraction (AVSE) aims to isolate a target speaker by leveraging both speech and complementary visual cues. This approach is inspired by the human perceptual ability to selectively attend to a single speaker’s voice through audio-visual integration in the brain. Within the broader family of speech enhancement and separation methods, audio-visual speaker extraction removes the need for pre-enrolled reference speech used in traditional speaker extraction, while also avoiding the permutation ambiguity problem common in speech separation. Since the target speaker’s speech is the primary focus of most speech processing applications, audio-visual speaker extraction serves as a fundamental front-end component for various downstream tasks, such as speaker diarization \cite{pan2024late}, speech recognition \cite{wang2024predict}, and speaker tracking 
\cite{qian2022audio}.

Despite the rapid progress in this field, relatively little attention has been paid to visual disturbances, which are common in real-world scenarios. Examples include lip occlusion caused by hands or microphones, video blur, and cases where the speaker moves out of the frame. Recent preliminary studies have begun exploring this direction by incorporating synthetically impaired visual data during training \cite{afouras2019my, sato2021multimodal, wu2022time, liu2024two, pan2024ravss}. In \cite{afouras2019my}, a two-stage pipeline is proposed: the first stage extracts the target speech using both audio and visual inputs, and the second stage enhances the speech using only the audio modality. In \cite{sato2021multimodal}, an additional enrollment speech is fused with visual cues to improve robustness. Under the speech separation setting, where all face tracks are available, visual degradation in one speaker’s video can be compensated by others \cite{wu2022time, liu2024two, pan2024ravss}. However, visual degradation in real-world scenarios can occur in diverse ways, and synthetic data cannot fully capture this variability. Therefore, it is preferable to develop mechanisms that do not rely on synthetically degraded visual data.

To address this challenge, a straightforward solution is to dig out different types of cues through audio-visual interactions, where the individually mined cues can effectively guide speaker extraction, and can compensate for each other when one of them is adversely affected. Since speaker embeddings are an effective cue for speaker extraction, several studies have attempted to derive them solely from the visual modality \cite{wang2022multi, gao2021visualvoice}. However, inferring speaker embeddings purely from face images is likely infeasible. Human perception studies show that people cannot reliably match an unfamiliar face to its corresponding voice (timbre) \cite{lavan2021explaining}. Li 
\etal separate speaker information from the visual modality and demonstrate that visual cues provide a useful signal only when the speakers differ in gender. In other words, models can capture identity-related features but not speaker embeddings from the visual modality alone. Therefore, speaker embeddings should be learned from audio-visual interactions rather than from visual information alone \cite{li2023rethinking}. 

Another commonly discussed cue in the literature is the synchronisation cue \cite{pan2021muse}. However, this concept is often defined too broadly, making it difficult to distinguish more specific types of synchronisation. The McGurk effect \cite{mcgurk1976hearing} provides insight into two distinct forms of synchronisation between the audio and visual modalities, namely acoustic synchronisation and semantic synchronisation. As demonstrated by Grant and Seitz \cite{grant2001effect}, the degree of mouth opening correlates with the F2-filtered speech signal, representing acoustic synchronisation. In contrast, semantic synchronisation refers to the alignment of speech and lip movements that correspond to the same linguistic content, such as speaking the same word simultaneously. The McGurk effect  shows that when humans are presented with mismatched audio-visual pairs that are acoustically synchronized but semantically misaligned, they often fail to detect the mismatch and instead perceive a fused or alternative word corresponding to either the audio or the visual input.

Building upon the above findings, we propose a robust audio-visual speaker extraction framework that incorporates a novel audio-visual learner and a cue interaction module. Specifically, the audio-visual learner leverages cross-modal interactions to progressively suppress interference in the speech signal and to hierarchically disentangle three key cues—speaker information, acoustic synchronisation, and semantic synchronisation. The cue interaction module then takes as input the speech representation and the three distinct cues. We design a dedicated interaction mechanism between the speech representation and each cue, allowing them to be processed separately. During this interaction, the speech representation is further enhanced, while the reliability of each cue is simultaneously estimated. Subsequently, an attention module fuses the three enhanced speech representations according to their respective reliabilities. Finally, a backend module reconstructs the clean speech signal from the fused representation.

Our contribution is summarized as follows:

\begin{itemize}
\item We address the degradation of audio-visual speaker extraction caused by impaired visual data by identifying and leveraging three distinct and effective cues—without relying on degraded visual data during training.
\item We propose an audio-visual learner that disentangles these three cues through hierarchical audio-visual interaction.
\item We design a cue interaction module that estimates the reliability of each cue and dynamically fuses them to enhance the speech representation.
\item Extensive experiments on the 
LRS3 and Voxceleb2 datasets demonstrate the effectiveness and robustness of the proposed framework under various types and levels of visual degradation.
\end{itemize}

\section{Related Work}

\subsection{Effective Cues in Speaker Extraction}
Speaker extraction can rely on several auxiliary cues that help locate or identify the target voice in a mixture \cite{zmolikova2023neural, xu2023multi}. Typical sources include speaker identity \cite{zhang2025multi,zeng2025usef}, semantic information \cite{kim2025contextual, ohishi2022conceptbeam, li2025sense}, and visual cues \cite{kalkhorani2025av, tan2020audio}.

\textbf{Speaker identity} is often obtained from a clean reference utterance of the target speaker. Such embeddings provide stable segment-level consistency but lack temporal synchrony with speech. Moreover, when interfering speakers have similar timbre or pitch, identity-based methods tend to fail.

\textbf{Semantic information} encompasses linguistic and contextual dependencies that indicate what content is likely spoken by the target. These cues provide high-level temporal consistency and can serve as indirect guidance when low-level acoustic or visual evidence is unreliable, offering a robust but less temporally aligned complement to other features.

\textbf{Visual cues} offer a richer and more dynamic modality. Lip motion aligns closely with acoustic energy and pitch variations, providing strong temporal synchronisation \cite{wang2022multi}, while facial appearance contributes weaker yet stable identity hints \cite{li2023rethinking}. However, their reliability depends heavily on video quality and visibility.

In addition, spatial information is also commonly employed as an auxiliary cue  \cite{quan2024spatialnet, luo2019fasnet, gu2020multi, michelsanti2021overview}. Although it may be affected by reverberation or overlapping sound sources, spatial cues are generally independent and stable, providing reliable localization of the target source. However, their reliance on multi-channel microphone arrays makes them not always available in practice.

Overall, these cues differ in both stability and temporal dynamics. Identity and contextual cues are relatively stable and global, while semantic cues exhibit stronger temporal variation and finer linguistic alignment. Visual cues, in contrast, are highly dynamic yet easily degraded—an imbalance that motivates the further analysis presented in the following sections.

\subsection{Visual Degradation}
In real-world scenarios, visual signals often suffer from degradation such as blur, occlusion, low resolution, lighting variation, or temporal desynchronisation \cite{wu2022time}. These corruptions make conventional visual encoders unreliable: the extracted embeddings may deviate from the true target cues or even fail to provide valid guidance. Consequently, the performance of AVSE methods drops sharply when video quality deteriorates.

To address the potential degradation of a single modality, several studies have introduced additional cues as complementary information. Some works introduce additional pre-enrolled speaker embeddings to complement degraded visual information~\cite{sato2021multimodal, wu2024unified}. However, these approaches rely on external auxiliary data and often fuse multiple cues through feature concatenation, multi-stage fusion, or cross-attention mechanisms, without explicitly evaluating the reliability of each cue. This design limits their generalization and flexibility in unconstrained conditions.

Other efforts attempt to model visual reliability within the AVSE model itself, typically by assigning a global confidence to the visual branch or by gating its contribution dynamically. Li et al.\
\cite{li2024momuse} maintain an internal speaker representation within the network and introduce a dual-path competitive mechanism between the internal representation and the current visual embedding, enabling the system to rely on the more reliable cues for target speech extraction. Furthermore, maintaining internal semantic representations can also serve as an auxiliary strategy to enhance network stability \cite{li2025memo}. While this mitigates gross visual corruption, such global reliability estimation treats the visual modality as a whole and fails to exploit the diverse, fine-grained information it carries, while still relying on degradation-augmented training data for robustness.

\subsection{Modeling Functional Visual Cues}
In this work, we revisit the structure of visual information and jointly analyze it with the mixed audio to uncover its intrinsic organization. Through this analysis, we disentangle the visual modality into three \textbf{functional cues}---speaker identity, acoustic synchronisation, and semantic synchronisation. This decomposition enables a more comprehensive utilization of visual information in AVSE, while offering complementary robustness across diverse environmental conditions. Each cue contributes differently yet cooperatively to the target extraction.

\textbf{Speaker identity} is not derived from a pre-enrolled reference, but implicitly inferred through global alignment between the video stream and the mixed audio. It provides an overall target indication with moderate temporal coherence, functioning as a soft identity constraint rather than a static enrollment.
\textbf{Acoustic synchronisation} reflects the direct correspondence between lip motion and acoustic features such as pitch and energy. It delivers the most immediate and fine-grained cues but is also the most sensitive to visual degradation.
\textbf{Semantic synchronisation}, analogous to visual speech recognition, attempts to infer high-level linguistic content from the video through an auxiliary semantic loss. Compared to acoustic cues, it exhibits stronger internal consistency and greater robustness to visual corruption, albeit at higher modeling difficulty.

Without using any pre-enrolled references or degraded videos during training, our approach achieves a principled decomposition of the visual modality into interpretable cues. This formulation enhances both the interpretability and robustness of AVSE models, significantly improving performance under adverse visual conditions.

\section{Methodology}

\label{sec:Method}
\begin{figure*}[!htb]
    \centering
    \includegraphics[width=1\textwidth]{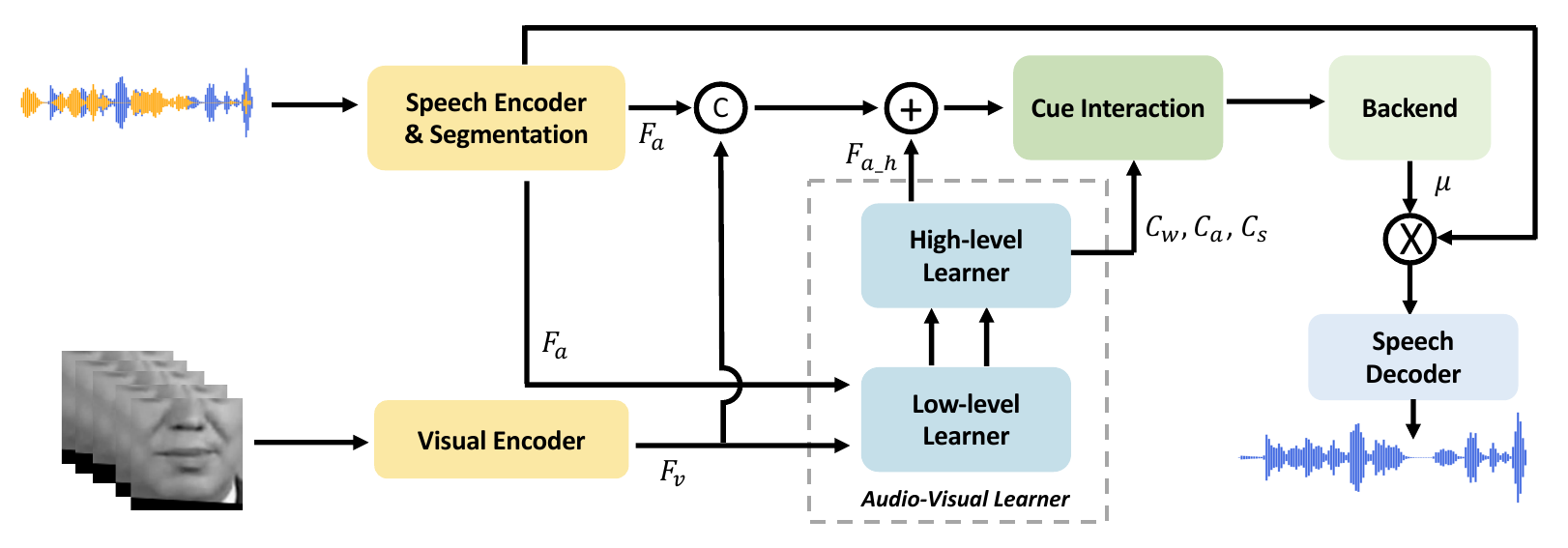}
    \caption{Pipeline of CueNet. The audio-visual learner performs cross-modal interaction between the audio feature $F_a$ and the lip-movement feature $F_v$, and hierarchically disentangles three cues: speaker, acoustic, and semantic cues. The cue interaction module then leverages these three cues to extract features belonging to the target speaker. Finally, the backend module reconstructs the target speech signal.
    }
    \label{fig:arch}
\end{figure*}

We propose CueNet, as illustrated in Fig. \ref{fig:arch}, which disentangles three distinct cues from audio-visual interactions and enables mutual compensation among them to enhance robustness against visual degradation. The overall architecture consists of a speech encoder, a visual encoder, an audio-visual learner, a cue interaction module, a backend module, and a speech decoder. Specifically, the backend module, the speech encoder, the decoder, and the visual encoder are introduced along with the task definition and formulation in Sec. \ref{sec:task}. The audio-visual learner, designed to disentangle the three cues through hierarchical interactive learning, is detailed in Sec. \ref{sec:learner}. The cue interaction module, which describes how CueNet dynamically fuses the three cues, is presented in Sec. \ref{sec:interaction}. Finally, the overall loss functions are summarized in Sec. \ref{sec:losses}.

\subsection{Task Formulation}
\label{sec:task}

We define the speech mixture as $X\in \mathbb{R}^{T_a}$, which is the sum of the target speech $S\in \mathbb{R}^{T_a}$ and the interfering speech $O\in \mathbb{R}^{T_a}$. Meanwhile, a video segment $V\in \mathbb{R}^{T_v}$ that is temporally synchronized with $S$ is available to complement the speech modality. Our objective is to estimate the target speech $\hat{S}$ from the mixture $M$ with the assistance of $V$, such that $\hat{S}$ closely approximates $S$ while effectively suppressing components of $O$.

\textit{The speech encoder} aims to downsample the mixture 
 $X$ and extract its feature representation $f_a\in \mathbb{R}^{N*T_f}$, where $N$ and $T_f$ denote the feature dimension and the feature length, respectively. Convolutional layers, real–imaginary spectrograms, and magnitude–phase spectrograms are commonly used as major encoder structures. Although real–imaginary spectrograms have recently demonstrated superior performance \cite{kalkhorani2025av, wang2023tf}, convolutional layers still provide greater flexibility for end-to-end, data-driven learning and enable lower latency during inference. Therefore, we use a 1D convolutional layer followed by a PReLU activation function as the encoder.
 
 Then, we segment $f_a$ along the length dimension using a sliding window with a stride equal to half the window length, obtaining $F_a$. By appropriately setting the window size and stride, $F_a$ has the shape $\mathbb{R}^{N*K*T_v}$. 

\textit{The visual encoder} extracts lip-movement features $F_v\in \mathbb{R}^{N*T_v}$ from face videos. Following most existing AVSE methods \cite{mu2024separate, zhao2025clearervoice, lin2023av}, we use ResNet-18 \cite{he2016deep} as the visual encoder. 

\textit{The backend module} estimates a mask $\mu$ which highlights the component of $S$ in $F_a$ while filtering out those belonging to $O$, given the output of the cue interaction module. Note that the visual modality is not involved in this module. We employ the backend of SeaNet \cite{tao2025audio} as our backend, which consists of multiple identical blocks. Each block contains a pair of inter-chunk and intra-chunk RNN modules for modeling the target speech, and another pair for modeling the interference speech. A cross-attention module is then applied to further separate them.

\textit{The speech decoder} reconstructs the target speech $\hat{S}$ from the masked speech feature $F_a'$, where $F_a'$ is obtained by element-wise multiplying $F_a$ with the estimated mask $\mu$. A linear layer is utilized to map $F_a'$ to speech waveforms.

\subsection{Audio-Visual Learner}
\label{sec:learner}

\begin{figure}[!htb]
    \centering
    \includegraphics[width=0.5\textwidth]{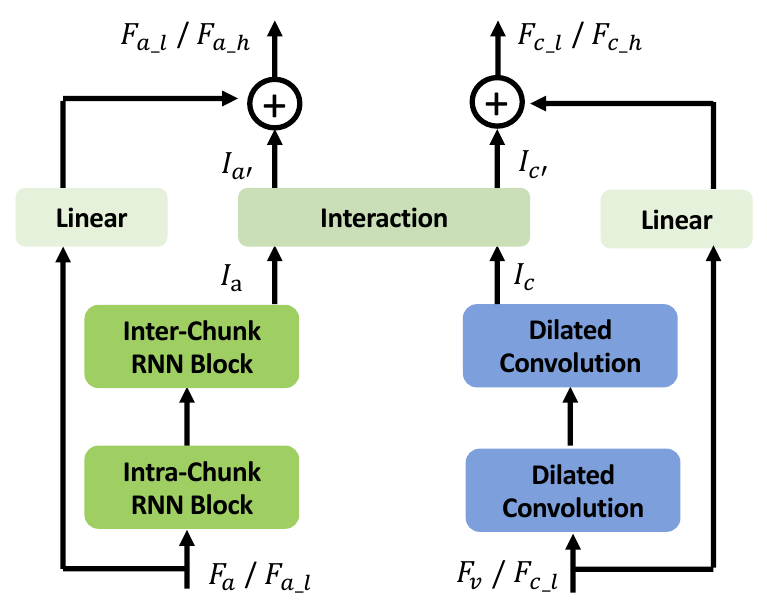}
    \caption{Low-level or high-level learners. If an entry contains a slash (/), the features before and after the slash correspond to those in the low-level and high-level learners, respectively.}
    \label{fig:learner}
\end{figure}

This module utilizes cross-modal interaction to achieve two goals. The main goal is to while gradually transforming video to three distinct cues: speaker embedding $C_s \in \mathbb{R}^{H}$, acoustic synchronisation $C_a\in \mathbb{R}^{H*T_v}$, and semantic synchronisation $C_w\in \mathbb{R}^{H*T_v}$, where $H$ is the number of hidden features. And another goal is coarsely detaching interference from audio. The two goals are compatible since both them only focus on the component of the target speaker. Meanwhile, as discussed in the introduction, speaker embeddings cannot be derived solely from the visual modality, whereas the audio and visual modalities complement each other in capturing acoustic and semantic synchronisation. 

Two sub-modules are hierarchically stacked to extract the three cues. The low-level learner focuses on extracting speaker information and acoustic synchronisation, while the high-level learner captures semantic synchronisation. As discussed in the introduction, acoustic synchronisation refers to the correlation between the degree of mouth opening and the acoustic spectrogram, particularly within the F2 frequency band. To model this relationship, we extract frame-wise pitch and spectrogram features from the target speech $S$, and cluster them using k-means into discrete tokens that serve as supervision signals, denoted as $Y_a$. Although pitch and energy vary dynamically during speech production \cite{rodero2017pitch}, they remain highly speaker-specific \cite{abberton1978intonation}. Therefore, the low-level learner jointly estimates acoustic synchronisation and speaker information, where the speaker ID is used as the supervision signal for the latter.

As illustrated in Fig. \ref{fig:learner}, $F_a$ and $F_v$ first pass through their respective inner-modality update modules. Since variants of combined inter-chunk and intra-chunk modeling have proven effective in speech processing \cite{luo2020dual, wu2019time, wu2019time}, we adopt a similar design and employ RNNs to capture both inter- and intra-chunk dependencies in the speech features, resulting in $I_a$. For the visual feature $F_{v}$, dilated convolutions are applied to capture long-range temporal dependencies. Subsequently, an interaction block is used to update the speech and visual features through cross-modal interaction, as formulated in Eq. \ref{eq:int_av} and Eq. \ref{eq:int_va}.

\begin{equation}\label{eq:int_av}
  I_{a'} = I_{a} + \mathbb{G}(I_{c} \cdot \sigma(\mathbb{P}(I_{a})))
\end{equation}

\begin{equation}\label{eq:int_va}
  I_{c'} = I_{c} + \mathbb{G}(I_{a} \cdot \sigma(\mathbb{P}(I_{c}))),
\end{equation}
where $\mathbb{G}$ and $\mathbb{P}$ denote networks, and $\sigma$ indicates the sigmoid function. 

Subsequently, two separate linear layers are applied to map $I_{c'}$ to the speaker information $C_s$ and acoustic synchronisation $C_a$, respectively, which are optimized using cross-entropy losses with their respective supervision signals. Since supervision for $C_s$ and $C_a$ may cause the loss of semantic information in $F_{c\_i}$, we incorporate the visual feature
 $F_v$ by passing it through a linear layer and adding it to $I_{c'}$, obtaining $F_{c\_l}$.

The high-level learner adopts the same architecture as the low-level learner, aiming to further purify the target speech while extracting semantics synchronisation from the interaction. Specifically, this module takes $F_{a\_l}$ and $F_{c\_l}$ as inputs and outputs $F_{a\_h}$ and $F_{c\_h}$. Unlike the low-level learner, which maps $I_c'$ to speaker information and acoustic synchronisation, the high-level learner maps $F_{c\_h}$ to the semantic synchronisation,  supervised by the tokens clustered via k-means from the features of the last encoder layer in AV-HuBERT \cite{shilearning}, denoted as $Y_w$.

$C_s$, $C_a$, and $C_w$ are optimized using cross-entropy losses with their individual supervision signals. Their corresponding losses are denoted as $\mathcal{L}_{spk}$, $\mathcal{L}_{acoustic}$, $\mathcal{L}_{semantics}$, respectively. We take $\mathcal{L}_{semantics}$ as an example, as defined in Eq. \ref{eq:ac}. 

\begin{equation}\label{eq:ac}
    \mathcal{L}_{semantics} = - Y_{w} \log (\mathbb{F}(C_w))
\end{equation}
where $\mathbb{F}$ is a linear layer.

\subsection{Cue Interaction Module}
\label{sec:interaction}
This module aims to effectively utilize the three estimated cues and achieve a dynamic fusion, where the weights of the cues are adaptively adjusted.

\begin{figure}[!htb]
    \centering
    \includegraphics[width=0.4\textwidth]{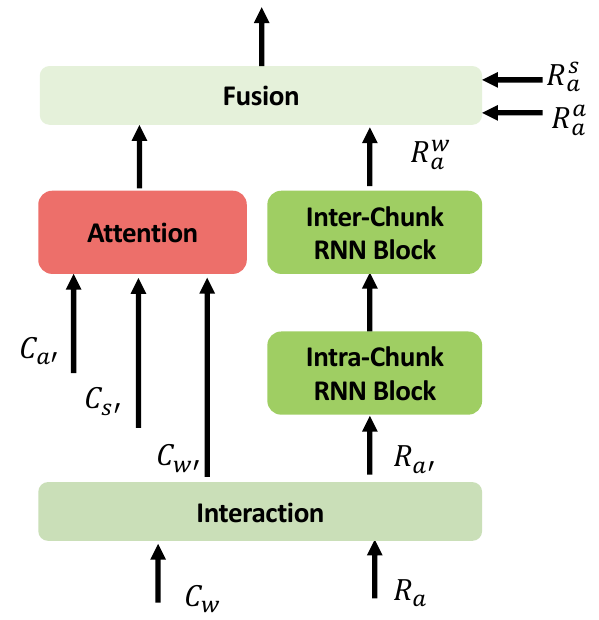}
    \caption{The Cue Interaction Module. Three cues $C_s$, $C_a$, and $C_w$, are individually applied to extract features belonging to the target speaker. The cue-enhanced features are then dynamically fused. $R_a$ is the speech feature, denotes the speech feature, computed by summing $F_a$ and $F_{a\_h}$ in Fig. \ref{fig:arch}.}
    \label{fig:cuei}
\end{figure}

For the sake of conciseness, we omit the interaction modules between $C_a$ and $R_a$, and between $C_s$ and $R_a$ in Fig. \ref{fig:cuei}, as their generation processes are identical to that between $C_w$ and $R_a$, which produces the semantic reliability $C_{w'}$ and the semantic-enhanced speech feature $R_{a'}$ shown in Fig. \ref{fig:cuei}. We compute $C_{w'}$ as defined in Eq. \ref{eq:real}. Meanwhile, through the interaction between $R_a$ and $C_w$, semantic-enhanced speech feature belonging to the target speaker is extracted into $R_{a'}$, as defined in Eq. \ref{eq:cue_feat}.

\begin{equation}\label{eq:real}
  C_{w'} = \mathbb{H}(C_{w}) \cdot \sigma(\mathbb{Q}(\mathcal{A}(R_a))),
\end{equation}

\begin{equation}\label{eq:cue_feat}
  R_{a'} = R_{a} + \mathbb{G}(C_{w} \cdot \sigma(\mathbb{P}(R_{a})))
\end{equation}
where $\mathbb{H}$, $\mathbb{Q}$, $\mathbb{G}$, $\mathbb{P}$ denote networks, and $\mathcal{A}$ is the average function on the dimension $K$ of the $R_a \in \mathbb{R}^{H*K*T_v}$, $H$ is the feature dimension. 

The attention module aims to dynamically determine the contributions of the three cues across both feature and temporal dimensions. Specifically, the three cues are stacked together to form a tensor with a shape of $J*H*T_v$, where $J$ is 3. Inspired by the concept of depthwise separable convolution \cite{szegedy2016rethinking}, where each convolutional kernel operates along a single dimension, we compute dependencies along the $T_v$ and $J$ dimensions separately using convolutional layers. A softmax function is then applied along the dimension of size $J$ to obtain the final attention weights, denoted as $\psi \in \mathbb{R}^{3*H*T_v}$.  

Given three pairs of cue reliability and cue-enhanced features, we fuse them to compute the refined speech feature $F_a^f$ corresponding to the target speaker. Particularly, we stack the three cue-enhanced speech features ($R_a^w$, $R_a^s$, $R_a^a$) into $R_a^f \in 3*H*K*T_v$. As $\psi$ represents the reliability of each cue for every frame and feature activation, we multiply it element-wise with $R_a^f$, then sum the result along the first dimension, yielding $R_e \in H*K*T_v$, which is subsequently fed into the backend module.

\subsection{Losses}
\label{sec:losses}

In addition to the aforementioned losses, we employ the SI-SNR objective \cite{le2019sdr} (as defined in Eq. \ref{eq:snr}) and multi-resolution spectrogram magnitude losses (as defined in Eq. \ref{eq:stft_loss}) to optimize the estimated target speech $\hat{S}$.

\begin{equation}\label{eq:snr}
  \mathcal{L}_{si\_snr} = 10 \log_{10} \frac{||\alpha S||^2}{||\hat{S} - \alpha S||^2}
\end{equation}
where $S$ and $\hat{S}$ denote the ground-truth target speech signal and the predicted target speech signal, respectively. $\alpha$ is a scalar that ensures scale invariance and is computed according to Eq. \ref{eq:scale}. 

\begin{equation}\label{eq:scale}
  \alpha = \frac{<\hat{S}, S>}{||S||^2}
\end{equation}

\begin{equation}\label{eq:stft_loss}
  \mathcal{L}_{stft} = \sum_{\theta}^{\Theta}|(||\Phi(\hat{S}, \theta)|| - ||\Phi(S, \theta)||)|,
\end{equation}
where $\Theta$ indicates the parameter set of the Short-Time Fourier Transform (STFT), $\Phi$ denotes the STFT function, $||\cdot||$ computes the spectrogram magnitude, and $|\cdot|$ represents the element-wise L1 loss. 

In summary, the training objective consists of five losses: $\mathcal{L}_{stft}$, $\mathcal{L}_{SI-SNR}$, $\mathcal{L}_{spk}$, $\mathcal{L}_{semantics}$, and $\mathcal{L}_{acoustic}$ during training. The last three are cross-entropy losses applied to the hidden representations within the audio-visual learner. All losses are combined with appropriate weighting to form the final optimization objective.

\section{Experiments}
\subsection{Experimental Configuration}

\begin{table*}[ht]
\centering
\small
\caption{Comparison with state-of-the-art methods on the LRS3 dataset under clean and degraded visual conditions. Four types of visual degradation are considered, where 50\% of the visual data is corrupted. GB: Gaussian Blur, CC: Concealment, MF: Masked Feature, and FM: Face Missing. Multiply–Accumulate operation (MAC) denotes the \textbf{computational cost} during inference. \textbf{\textit{Values}} in bold italics indicate the \textbf{best} performance, while \textbf{values} in bold indicate the \textbf{second-best} performance. SI-SNRi and SDR are reported in dB.} 
\setlength{\tabcolsep}{4pt} 
\begin{tabular}{llccccccccccccccc}
\toprule
\multirow{2}{*}{\textbf{Method}} &
\multirow{2}{*}{\textbf{MAC}} &
\multicolumn{5}{c}{\textbf{SI-SNRi}} &
\multicolumn{5}{c}{\textbf{SDR}} &
\multicolumn{5}{c}{\textbf{PESQ}} \\
\cmidrule(lr){3-7} \cmidrule(lr){8-12} \cmidrule(lr){13-17}
 &  &
 \textbf{Clean} & \textbf{GB} & \textbf{CC} & \textbf{MF} & \textbf{FM} &
 \textbf{Clean} & \textbf{GB} & \textbf{CC} & \textbf{MF} & \textbf{FM} &
 \textbf{Clean} & \textbf{GB} & \textbf{CC} & \textbf{MF} & \textbf{FM} \\
\midrule
VisualVoice \cite{gao2021visualvoice} & 16.13 & 9.9 & - & - & - & - & 10.3 & - & - & - & - & 2.13 & - & - & - & - \\
ConvTasNet \cite{wu2019time} & 38.80 & 11.2 & - & - & - & - & 11.7 & - & -  & - & - & 2.58 & - & - & - & \\
AVLiT-8 \cite{martel23_interspeech} & 35.52  & 14.3 & 12.3 & 12.9 & 12.4 & 11.8 & 14.5 & 12.8 & 13.2 & 12.7 & 12.2 & 2.44 & 2.19 & 2.27 & 2.23 & 2.17 \\
CTCNet \cite{li2024audio} & 255.6  & 17.4 & 16.0 & 14.7 & 15.0 & 13.7 & 17.6 & 16.5 & 15.5 & 15.7 & 14.8 & 2.90 & 2.80 & 2.74 & 2.74 & 2.68\\
Seanet \cite{tao2025audio} & 18.76 & 16.8 & 15.3 & 15.1 & 14.1 & 13.6 & 16.9 & 15.8 & 15.4  & 14.8 & 14.0 & 3.09 & 2.82 & 2.78 & 2.74  & 2.69\\
IIANet-fast \cite{li2024iianet} & 17.83 & 17.6 & 15.4 & 14.3 & 15.7 & 12.1 & 17.8 & 16.3 & 15.5 & 16.4 & 12.7  & 2.91 & 2.79 & 2.73 & 2.80 & 2.51  \\
IIANet \cite{li2024iianet} & 35.30 & \textbf{18.4} & 15.8 & 14.7 & 16.1 & 12.8 & \textbf{\textit{18.6}} & 16.5 & 15.7 & 16.6 & 13.5 & 3.09 & 2.89 & 2.83 & 2.87 & 2.58\\
\midrule
CueNet-fast & 15.12 & 17.9 & \textbf{16.8} & \textbf{17.4} & \textbf{16.6} & \textbf{15.9} & 18.0 & \textbf{17.1} & \textbf{17.5} & \textbf{16.8} & \textbf{16.3} & \textbf{3.14} & \textbf{2.96} & \textbf{3.02} & \textbf{2.92} & \textbf{2.88}\\
CueNet & 24.77 & \textbf{\textit{18.5}} & \textbf{\textit{17.3}} & \textbf{\textit{17.7}} & \textbf{\textit{16.9}} & \textbf{\textit{16.7}} & \textbf{\textit{18.6}} & \textbf{\textit{17.6}} & \textbf{\textit{17.9}} & \textbf{\textit{17.1}} & \textbf{\textit{17.0}} & \textbf{\textit{3.24}} & \textbf{\textit{3.05}} & \textbf{\textit{3.10}} & \textbf{\textit{3.00}} & \textbf{\textit{3.00}}\\

\bottomrule
\end{tabular}
\label{tab:lrs3}
\end{table*}

\hspace{1em} \textbf{\textit{Dataset:}} We conduct experiments on two widely used datasets: LRS3 \cite{afouras2018deep} and VoxCeleb2 \cite{chung2018voxceleb2}. LRS3 is an audio-visual speech recognition dataset compiled from TED Talks, providing \textbf{high-quality audio recordings} with minimal background noise. In contrast, VoxCeleb2 is designed for audio-visual speaker recognition and is collected from in-the-wild interview videos, where \textbf{background noise} of varying intensity is present. Following the configurations in \cite{li2024iianet, lee2021looking}, each audio clip is cropped to a fixed duration of 2 seconds and resampled to 16\,kHz. The signal-to-noise ratio (SNR) between the target and interference speech varies from –5\,dB to 5\,dB. Face videos are cropped into 88×88 grayscale frames with the lips centered, and each video is recorded at 25\,FPS. Regarding dataset sizes, LRS3 consists of 28 hours of training data, 3 hours of validation data, and 1.5 hours of test data. VoxCeleb2 differs only in the amount of training data, which is expanded to 56 hours due to the presence of background noise.

\begin{table*}[ht]
\centering
\small
\caption{Comparison with state-of-the-art methods on the Voxceleb2 dataset under clean and degraded visual conditions. Four types of visual degradation are considered, where 50\% of the visual data is corrupted. GB: Gaussian Blur, CC: Concealment, MF: Masked Feature, and FM: Face Missing. Multiply–Accumulate operation (MAC) denotes the \textbf{computational cost} during inference. \textbf{\textit{Values}} in bold italics indicate the \textbf{best} performance. SI-SNRi and SDR are reported in dB.} 
\setlength{\tabcolsep}{4pt} 
\begin{tabular}{llccccccccccccccc}
\toprule
\multirow{2}{*}{\textbf{Method}} &
\multirow{2}{*}{\textbf{MAC}} &
\multicolumn{5}{c}{\textbf{SI-SNRi}} &
\multicolumn{5}{c}{\textbf{SDR}} &
\multicolumn{5}{c}{\textbf{PESQ}} \\
\cmidrule(lr){3-7} \cmidrule(lr){8-12} \cmidrule(lr){13-17}
 &  &
 \textbf{Clean} & \textbf{GB} & \textbf{CC} & \textbf{MF} & \textbf{FM} &
 \textbf{Clean} & \textbf{GB} & \textbf{CC} & \textbf{MF} & \textbf{FM} &
 \textbf{Clean} & \textbf{GB} & \textbf{CC} & \textbf{MF} & \textbf{FM} \\
\midrule
VisualVoice \cite{gao2021visualvoice}  & 16.13 & 9.3 & - & - & - & - & 10.2 & - & - & - & - & 2.45 & - & - & - & - \\
ConvTasNet \cite{wu2019time} & 38.80 & 9.2 & - & - & - & - &  9.8 & - & - & - & - &  2.17 & - & - & - & -  \\
AV-DPRNN \cite{zhao2025clearervoice} & - & 11.2 & - & - & - & - &  11.7 & - & - & - & - & 2.43 & - & - & - & - \\
Muse \cite{pan2021muse} & - & 11.7 & - & - & - & - & 12.0 & - & - & - & - & - & - & - & - & - \\
AV-TFGridNet \cite{zhao2025clearervoice} & - & 13.5 & 9.1 & 9.4 & 10.3 & 8.1 & 13.9 & 9.7 & 10.4 & 10.6 & 8.3 & 2.57 & 2.38 & 2.43 & \textbf{\textit{2.45}} & 2.35\\
CTCNet \cite{li2024audio} & 255.6 & 12.0 & 8.5 & 9.1  & 10.2 & 7.8 & 12.7 & 9.1 & 9.7 & 10.6 & 9.0 & 2.34 & 2.15 & 2.22 & 2.25 & 2.14 \\
AV-Sep \cite{lin2023av} & - & 12.1 & - & - & - & - &  - & - & - & - & - & 2.31 & - & - & - & -  \\
Seanet \cite{tao2025audio} & 18.76 & 13.2 & 9.5 & 7.0 & 8.9 & 9.7 & 13.6  & 10.7 & 8.9 & 10.3 & 10.9 & 2.58 & 2.38 & 2.31 & 2.37 & 2.38 \\
IIANet \cite{li2024iianet} & 35.30 & 13.6 & 9.8 & 10.3 & 10.5 & 9.8 & \textbf{\textit{14.3}}  & 10.9 & 11.3 & 11.3 & 10.9 & 2.61  & 2.33 & 2.36 &  2.38 & 2.33\\
IIANet-fast \cite{li2024iianet} & 17.83 & 12.6 & - & - & - & - & 13.6 & - & - & - & - & - & - & - & - & -  \\
\midrule
CueNet-fast & 15.12 & \textbf{\textit{13.7}} & \textbf{\textit{10.5}} & \textbf{\textit{11.5}} & \textbf{\textit{11.1}} & \textbf{\textit{10.6}} & 14.1 & \textbf{\textit{11.5}} & \textbf{\textit{12.4}} & \textbf{\textit{12.0}} & \textbf{\textit{11.7}} & \textbf{\textit{2.62}} & \textbf{\textit{2.38}} & \textbf{\textit{2.49}}  & 2.40 & \textbf{\textit{2.46}} \\

\bottomrule
\end{tabular}
\label{tab:vox2}
\end{table*}

\textbf{\textit{Metrics}} To evaluate the performance of our audio-visual speech extraction system, we adopt three widely used metrics: Scale-Invariant Signal-to-Noise Ratio improvement (SI-SNRi), Signal-to-Distortion Ratio (SDR), and Perceptual Evaluation of Speech Quality (PESQ). SI-SNRi measures the improvement in the scale-invariant SNR relative to the mixture. SDR provides a holistic assessment of reconstruction fidelity by quantifying the overall distortion between the estimated and reference signals. Complementing these objective distortion metrics, PESQ serves as a perceptual quality measure that estimates human-judged speech quality by modeling auditory perception.

\textbf{\textit{Visual degradation:}} We introduce four types of visual degradation during \textbf{inference} to simulate real-world visual disturbances. \textbf{\textit{Gaussian Blur (GB)}} commonly occurs when the camera or the speaker moves rapidly, or when the camera is out of focus. \textbf{\textit{Concealment (CC)}} represents scenarios where objects such as microphones or hands occlude the speaker’s lips. Following common practice in audio-visual processing under occlusion \cite{hong2023watch, wang2024restoring}, we use objects from the Naturalistic Occlusion Generation dataset \cite{voo2022delving} to simulate occlusions. \textbf{\textit{Masked Feature (MF)}} simulates a more challenging case where frames of visual features are missing. \textbf{\textit{Face Missing (FM)}} models situations where the speaker moves out of the camera frame.

A designated proportion of frames is subjected to each type of visual degradation. Instead of randomly masking individual frames, we follow the practice of many self-supervised methods and mask blocks of 5 consecutive frames \cite{baevski2022data2vec, shilearning}, as adjacent visual frames often convey similar information.

\textbf{\textit{Implementation details:}} Our models are trained on 4 GeForce RTX 3090 GPUs with an initial learning rate of 0.001. We use the Adam optimizer \cite{kingma2014adam} for training. Batch size is set as 14 per GPU. The ReduceLROnPlateau scheduler is utilized for better convergence. Specifically, the learning rate is reduced by half if the validation loss does not improve for 10 consecutive epochs. 

The hyperparameters are listed below. Three sets of STFT parameters are used in $\mathcal{L}_{stft}$. In particular, we adopt window and hop sizes of 60\,ms / 40\,ms, 40\,ms / 20\,ms, and 20\,ms / 10\,ms. For the speech encoder, the kernel size and stride are set to 40 and 20, respectively, and the output feature dimension is 256. We then apply a sliding window with a size of 64 and a stride of 32 to segment the features. The inter-chunk RNN block and the intra-chunk RNN block are implemented using bidirectional LSTMs \cite{graves2012long} with a hidden dimension of 128, followed by Group Normalization with a group size of 1. $\mathbb{G}$ and $\mathbb{P}$ in Eq. \ref{eq:int_va} are implemented as 1D convolutional layers followed by a global normalization layer. In the Audio-Visual Learner, each dilated convolution block consists of two 1D convolutional layers, with PReLU activation and layer normalization applied between them. The second 1D convolutional layer uses a dilation factor of 2 to model longer-range temporal dependencies. The hidden feature dimension is also set to 128.

As described in Sec. \ref{sec:task}, the backend module consists of multiple identical blocks. Following IIANet \cite{li2024iianet}, we implement two versions of CueNet with different computational complexities in terms of Multiply–Accumulate operations (MACs). Specifically, \textbf{CueNet-fast} employs 5 blocks in the backend module, whereas \textbf{CueNet} uses 10 blocks.

\begin{table}[ht]
\centering
\small
\caption{Comparison on LRS3 test set using models trained on \textbf{Voxceleb2}. SI-SNRi is reported in dB.} 
\setlength{\tabcolsep}{4pt} 
\begin{tabular}{lllccccc}
\toprule
\textbf{Metrics} & \textbf{Method} & \textbf{MACs} & \textbf{Clean} & \textbf{GB} & \textbf{CC} & \textbf{MF} & \textbf{FM}  \\
\midrule
\multirow{2}{*}{SI-SNRi} 
& IIANet  & 35.30 & 16.2 & 9.5 & 10.8 & 11.5 & 8.1 \\
& CueNet-fast & 15.12 & \textbf{16.7} & \textbf{12.8} & \textbf{15.2} & \textbf{14.6} & \textbf{13.9} \\
\midrule
\multirow{2}{*}{PESQ} 
& IIANet  & 35.30 & 2.79 & 2.52 & 2.53 & 2.56 & 2.43 \\
& CueNet-fast & 15.12 & \textbf{2.88} & \textbf{2.67} & \textbf{2.77} & \textbf{2.75} & \textbf{2.71} \\
\bottomrule
\end{tabular}
\label{tab:cross}
\end{table}

\subsection{Comparison of Experiment Result}

\hspace{1em} \textbf{\textit{Extraction performance on LRS3:}} As shown in Tab. \ref{tab:lrs3}, the proposed CueNet-fast has outperformed all baseline methods across all four types of visual degradation. In particular, the improvement under the face missing condition is the most pronounced. Since this degradation is the most challenging among all cases, the substantial gain demonstrates the robustness and advantage of our method. Although CueNet-fast performs slightly worse than IIANet in SI-SNRi and SDR when clean visual data is provided, its computational cost is considerably 
lower. When comparing CueNet-fast with IIANet-fast, which has a similar computational cost, CueNet-fast surpasses IIANet-fast on all metrics. Moreover, CueNet-fast achieves higher PESQ scores than IIANet, indicating that it preserves speech intelligibility more effectively. From another perspective, CueNet-fast is inferior to IIANet under clean visual conditions but surpasses it once visual degradation is introduced, indicating substantially better robustness. 

When comparing CueNet with IIANet, CueNet also outperforms IIANet under clean visual conditions. Meanwhile, CueNet still requires lower computational cost, which further confirms the superiority of the proposed method.

\begin{table}[ht]
\centering
\small
\caption{Ablation studies on the contribution of different cues and combination on the LRS3 dataset. Spk refers to speaker information. No vision degradation is applied.}
\begin{tabular}{ccc|ccc}
\hline
Spk & Acoustic & Semantic & SI-SNRi & SDR & PESQ \\
\hline
\checkmark & \checkmark & \checkmark & 17.9 & 18.0 & 3.14 \\
\checkmark & \checkmark  & $\times$ & 17.8 & 17.9 & 3.11 \\
\checkmark  & $\times$  & \checkmark & 17.6 & 17.7 & 3.06 \\
$\times$  & \checkmark & \checkmark  & 17.7 & 17.9 & 3.08 \\
\checkmark  & $\times$  & $\times$ & 17.4 & 17.6 & 3.04 \\
$\times$  & \checkmark  & $\times$ & 17.3 & 17.4 & 3.04 \\
$\times$  & $\times$  & \checkmark & 17.1 & 17.3 & 3.01 \\
$\times$  & $\times$  & $\times$ & 17.1 & 17.2 & 3.01 \\
\hline
\end{tabular}
\label{tab:ab}
\end{table}

\textbf{\textit{Extraction performance on Voxceleb2}}
Models trained on VoxCeleb2 exhibit noticeably lower performance compared to those trained on LRS3, primarily due to the presence of background noise in the VoxCeleb2 recordings. In addition, the performance degradation caused by impaired visual inputs is more pronounced on this dataset, as the visual modality plays a crucial role in guiding speaker extraction. Nevertheless, our model consistently achieves over 10\,dB SI-SNRi across all types of visual degradation. Moreover, it outperforms all baseline methods under every degradation setting. Although the SDR is slightly lower than that of IIANet, our SI-SNRi scores are higher, further confirming the advantages of the proposed approach.

\textbf{\textit{Cross-domain performance}} We evaluate the cross-domain generalization capability by training CueNet on Voxceleb2 and testing it on the LRS3 test set. We compare our model with the most competitive baseline, IIANet, and report the results in Tab.~\ref{tab:cross}.

From the SI-SNRi results, CueNet-fast consistently outperforms IIANet across all acoustic conditions. Specifically, our method surpasses IIANet by 0.5 dB, 3.3 dB, 4.4 dB, 3.1 dB, and 5.8 dB under Clean, GB, CC, MF, and FM conditions, respectively. Notably, these improvements are substantially larger than those observed in the in-domain setting (i.e., training and testing on Voxceleb2), where the corresponding gains are 0.1 dB, 0.7 dB, 1.2 dB, 0.6 dB, and 0.8 dB. This comparison further highlights the superior cross-domain robustness of our approach.

A similar trend is observed for PESQ, where CueNet-fast achieves consistent perceptual quality improvements across all acoustic conditions. These results demonstrate that our model not only improves signal-level separation performance but also enhances perceptual speech quality under domain shift.

Overall, the results confirm that our method exhibits stronger robustness and generalization ability when evaluated under unseen acoustic conditions.

\begin{figure*}[t!]
    \centering
    \begin{subfigure}[b]{0.45\textwidth}
        \centering
        \includegraphics[width=\textwidth]{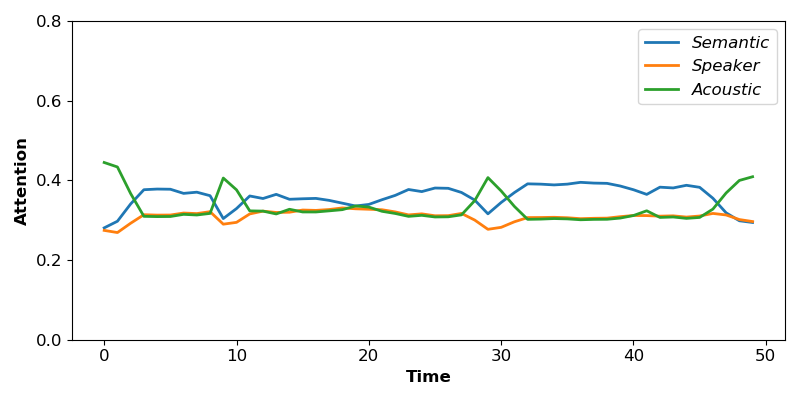}
    \end{subfigure}
    \hfill
    \begin{subfigure}[b]{0.45\textwidth}
        \centering
        \includegraphics[width=\textwidth]{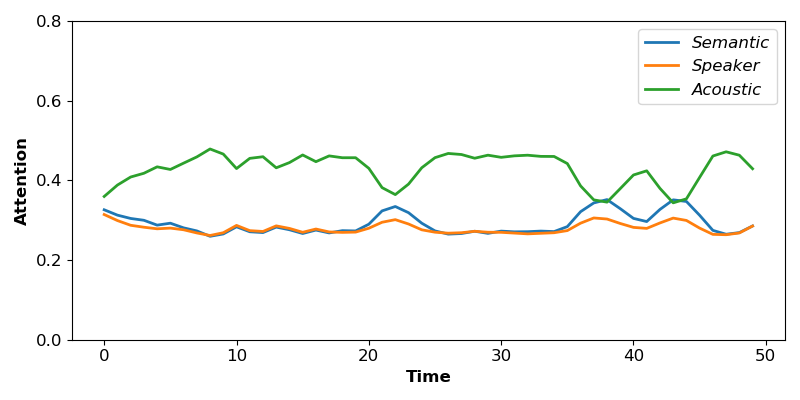}
    \end{subfigure}

    \vspace{0.5cm}

    \begin{subfigure}[b]{0.45\textwidth}
        \centering
        \includegraphics[width=\textwidth]{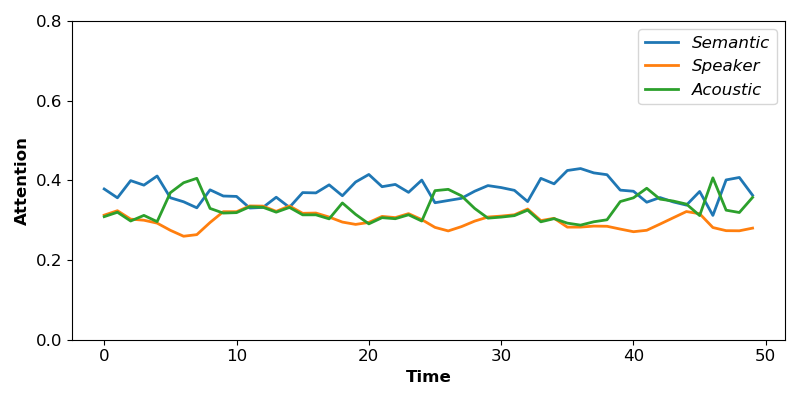}
    \end{subfigure}
    \hfill
    \begin{subfigure}[b]{0.45\textwidth}
        \centering
        \includegraphics[width=\textwidth]{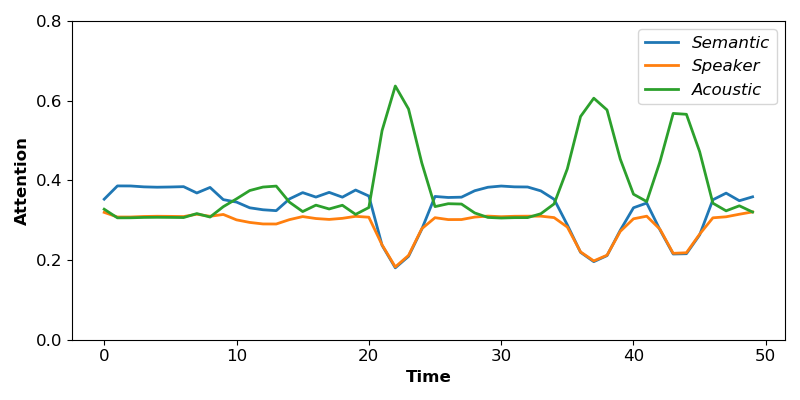}
    \end{subfigure}

    \caption{Attention cross the temporal dimension. There are 50 temporal frames.}
    \label{fig:att_value}
\end{figure*}

\subsection{Ablation Studies}

\textbf{\textit{Contribution of different cues}}
This section aims to validate the effectiveness of cue disentanglement, using \textbf{CueNet-fast}. We first examine the performance with clean video, as shown in Tab. \ref{tab:ab}.

The full model, which incorporates all three cues, achieves the best performance across all metrics (17.9\,dB SI-SNRi, 18.0\,dB SDR, and 3.14 PESQ). Removing any single cue leads to a performance drop, confirming that each cue contributes positively to the overall extraction quality.

Among the three cues, speaker information and acoustic synchronisation appear to be the most influential. When either of them is removed (second and third rows), the performance decreases noticeably, with larger drops observed when speaker information is excluded. In contrast, removing semantic synchronisation results in a smaller degradation, suggesting that while semantic alignment contributes to performance, its impact is comparatively weaker under clean visual conditions.

When only one cue is preserved (rows 5–7), the trend becomes even clearer: using speaker information alone performs best among single-cue settings, followed by acoustic synchronisation, and finally semantic synchronisation. The model without any cue guidance (last row) performs the worst, demonstrating that the cues extracted via audio-visual interaction are essential for effective speaker extraction.

\begin{figure}[t!]
    \centering
    \includegraphics[width=0.48\textwidth]{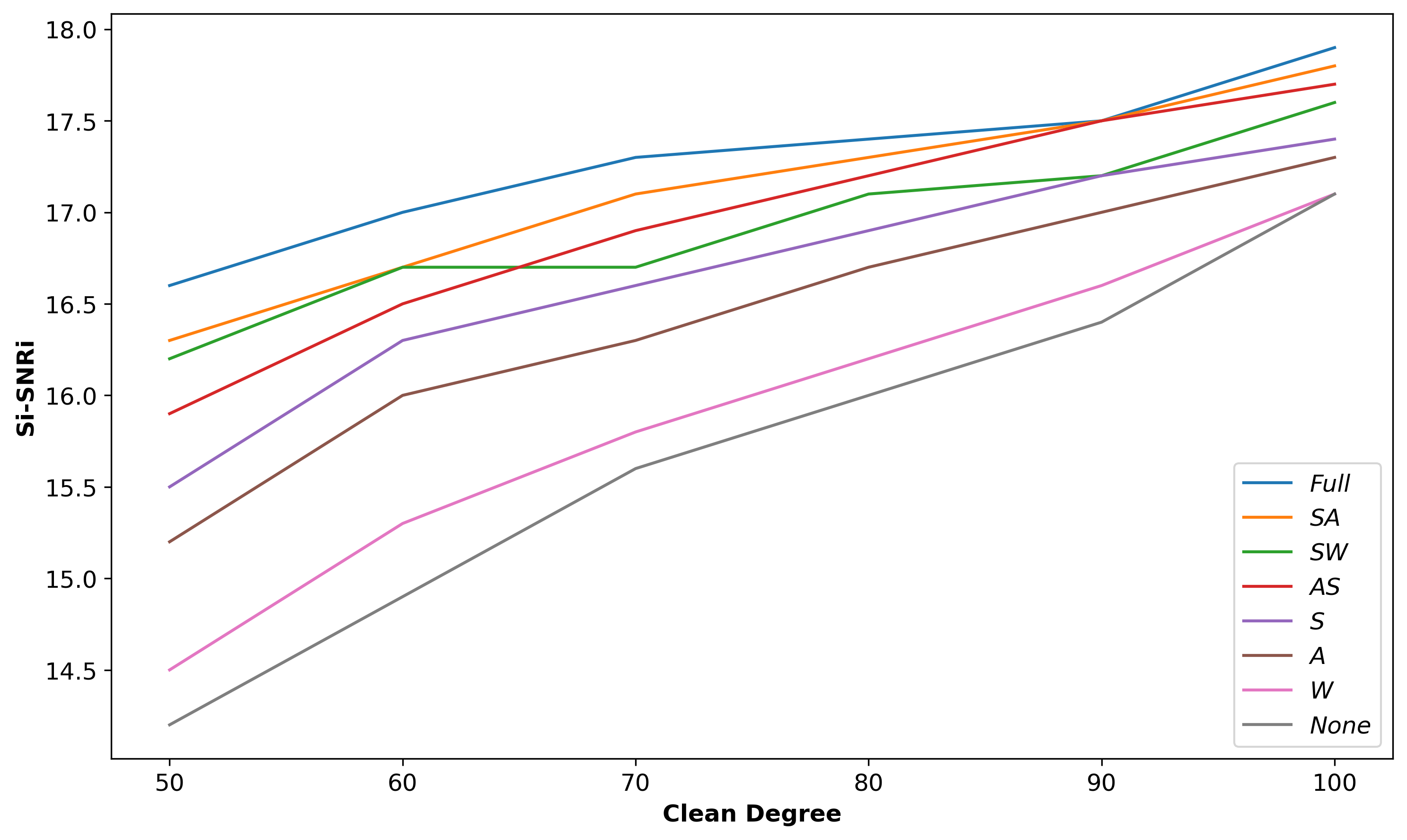}
    \caption{Performance across different degree of corrpution. Maksed feature is applied to simulate visual degradation. "A", "S", "W" indicate acoustic, speaker and semantic.
    }
    \label{fig:tren}
\end{figure}

We further examine the performance trend as visual degradation is gradually increased. As shown in Fig. \ref{fig:tren}, the performance gap between models that utilize more cues and those with fewer cues widens as the degradation severity grows. In addition, the decline in performance is relatively small at mild degradation levels but accelerates as the degradation becomes more severe. This pattern highlights the importance of each cue and the increasing benefit of incorporating all three when visual reliability deteriorates.

\begin{table}[h]
\centering
\small
\caption{Ablation studies on the cue interaction module on the LRS3 dataset.}
\label{tab:1}
\begin{tabular}{c|cc|cc}
\hline
Method & \multicolumn{2}{c|}{Clean} & \multicolumn{2}{c}{Masked Feature} \\
           & SI-SNRi  & SDR & SI-SNRi  & SDR \\
\hline
Cue Interation     & 17.9  & 18.0   & 16.6   & 16.8      \\
Concatenation     & 17.4   & 17.5   & 15.8   & 16.1      \\
\hline
\end{tabular}
\end{table}

\begin{table}[ht]
\centering
\small
\caption{Ablation studies on K-means using CueNet-fast. The metric is SI-SNRi, which is reported in dB.} 

\begin{tabular}{lcccc}
\toprule
Classes & 64 & 128 & 256 & 512  \\
\midrule
Clean Video & \textbf{17.94} & 17.93  & 17.89  & 17.85  \\ 
Masked Frame & 16.26 & 16.51  & \textbf{16.65}  & 16.59 \\
\bottomrule
\end{tabular}
\label{tab:kmean}

\end{table}

\textbf{\textit{Contribution of the cue interaction module}} This section evaluates the contribution of the proposed cue interaction module. We adopt simple concatenation as a baseline to assess whether dynamic cue fusion offers advantages beyond straightforward feature merging. Among the four types of visual degradation, we select the masked feature setting for this analysis because it provides fine-grained controllability over the amount and location of removed visual information.

As shown in Tab. \ref{tab:1}, the cue interaction module consistently outperforms concatenation in both clean and degraded conditions. Under clean visual inputs, the proposed interaction mechanism yields a clear improvement (17.9 vs.\ 17.4\,dB SI-SNRi), demonstrating that the dynamic weighting of cues is beneficial even when the visual modality is reliable. The performance gap becomes more pronounced under the masked feature scenario, where visual information is partially missing. In this case, cue interaction achieves 16.6\,dB SI-SNRi compared to 15.8\,dB from concatenation, confirming that adaptively modulating cue contributions is especially important when visual degradation occurs.

Overall, these results validate that the cue interaction module not only enhances performance in ideal conditions but also considerably 
improves robustness against impaired visual inputs, aligning with our core motivation.

\textbf{\textit{Impact of Cluster Number}} As we obtain supervision signals for the acoustic and semantic cues via K-means clustering, model performance may be influenced by the choice of $K$. For the semantic cue, we adopt 48 clusters to roughly align with the 44 phonemes in English \cite{bizzocchi2017many}, allowing a small margin to account for pronunciation variability.

For the acoustic cue, we use 512 clusters in Tab. \ref{tab:lrs3}–\ref{tab:vox2}, following a common practice in self-supervised learning frameworks that employ K-means clustering. However, this choice is empirical rather than theoretically grounded. Therefore, we conduct an ablation study on the impact of $K$ for the acoustic cue, as reported in Tab. \ref{tab:kmean}.

 Overall, the choice of $K$ does not have a huge impact. In detail, a smaller $K$ slightly favors performance under clean visual conditions, whereas a larger $K$ improves robustness when visual frames are masked. We hypothesize that, with clean video, audio–visual synchronisation are easier to exploit and overly fine-grained acoustic cues may introduce unnecessary noise. In contrast, when visual information is corrupted, a larger number of acoustic clusters provides more discriminative cues, leading to improved robustness.

\subsection{Cue-Attention Visualization}

Consistent with our motivation that the three cue-enhanced speech features are fused using dynamic weights to achieve better robustness, we visualize these weights across time. Since the weights have a shape of $H*T*3$ (where $H$ is the feature dimension), we randomly select and plot one feature of shape $T*3$ in Fig. \ref{fig:att_value}. Additional feature plots can be found in the Supplementary Material.

These visualizations highlight that the learned attention distribution is highly dimension-dependent. Some dimensions consistently assign high weight to acoustic cues, indicating that they function as acoustic specialists sensitive to energy variations, spectral transitions, or prosodic changes. Other dimensions rely predominantly on semantic information, suggesting these dimensions encode linguistically driven patterns useful for guiding speaker extraction. Additional dimensions exhibit a more balanced and dynamic allocation of attention, responding adaptively to cross-modal evidence. The diversity of behaviors across dimensions demonstrates that the model learns a rich, distributed strategy for cue fusion, in which different parts of the feature space specialize in different modalities. This specialization allows the model to robustly combine heterogeneous information sources when isolating the target speaker.

\section{Conclusion}

In this paper, we proposed a novel audio-visual speaker extraction framework that remains robust under visual degradation without requiring any impaired visual data during training. Motivated by insights from human perception, our audio-visual learner disentangles three complementary cues (speaker information, acoustic synchronisation, and semantic synchronisation) through hierarchical cross-modal interaction. We further introduced a dedicated cue interaction module that dynamically measures the reliability of each cue and effectively fuses the resulting cue-enhanced speech representations to guide extraction. Extensive experiments demonstrate that the proposed framework consistently outperforms state-of-the-art methods across various types of visual degradation, while ablation studies validate the individual contributions of the three cues and the effectiveness of the proposed fusion mechanism. These results confirm the strong robustness and practical potential of the proposed approach for real-world audio-visual speaker extraction.
Given the demonstrated robustness of our disentanglement mechanism, it will be of interest to investigate its potential benefits in other audio-visual tasks, such as speech recognition and active speaker detection.

\section{Acknowledgement}
\noindent This work was supported by the EU H2020 project No. 101135556 (INDUX-R), the DFG’s Reinhart Koselleck Project 442218748 (AUDI0NOMOUS).

\ifCLASSOPTIONcaptionsoff
  \newpage
\fi

\bibliographystyle{IEEEtran}
\bibliography{main}

\end{document}